Hindawi

## Research Article

# Dissociation of Quarkonium in Hot and Dense Media in an Anisotropic Plasma in the Nonrelativistic Quark Model

**M. Abu-Shady** ,[1] **H. M. Mansour,**[2] and **A. I. Ahmadov**[3]

[1]*Department of Applied Mathematics, Faculty of Science, Menoufia University, Egypt*
[2]*Department of Physics, Faculty of Science, Cairo University, Egypt*
[3]*Institute for Physical Problems, Baku State University, Z. Khalilov St. 23, AZ-1148 Baku, Azerbaijan*

Correspondence should be addressed to M. Abu-Shady; dr.abushady@gmail.com

In this paper, quarkonium dissociation is investigated in an anisotropic plasma in the hot and dense media. For that purpose, the multidimensional Schrödinger equation is solved analytically by Nikiforov-Uvarov (NU) method for the real part of the potential in an anisotropic medium. The binding energy and dissociation temperature are calculated. In comparison with an isotropic medium, the binding energy of quarkonium is enhanced in the presence of an anisotropic medium. The present results show that the dissociation temperature increases with increasing anisotropic parameter for 1S state of the charmonium and bottomonium. We observe that the lower baryonic chemical potential has small effect in both isotropic and anisotropic media. A comparison is presented with other pervious theoretical works.

## 1. Introduction

The dissociation of heavy quarkonium has been suggested in the formation of quark-gluon plasma (QGP) in the pioneering work of Matsui and Sats [1]. The production or suppression of quarkonia has been investigated either theoretically or experimentally in Refs. [2–12], and the disassociation temperature has been studied in Refs. [13–17]. In addition, ultrarelativistic heavy-ion collisions (URHIC) are used to explore the quark-gluon plasma (QGP) media as in [18–20]. This shows that the present topic is an interesting research work from both theoretical and experimental views.

One of the most remarkable features of the QGP formation is the color screening of the static chromoelectric fields [21]. The Debye screening in QCD plasma has been studied as a probe of deconfinement in a dense partonic medium which shows a reduction in the interaction between heavy quarks and antiquarks due to color screening leading to a suppression in $J/\Psi$ yields [22, 23]. Hence, the quarkonium in a hot medium is a good tool to examine the confined/deconfined state of matter.

Recently, the quarkonium properties have been studied by modifying both the Coulombic and string terms of the heavy-quark potential using the perturbative hard thermal loop (HTL) dielectric permittivity in both the isotropic and anisotropic media in the static [24–28] and in the moving media [29]. Many attempts have been suggested to calculate the dissociation temperatures of quarkonium states in the deconfined medium by applying the lattice calculations of quarkonium spectral functions [30–33] or nonrelativistic calculations which depend on some effective (screened) potentials [34–39] in an isotropic medium. The nuclei moving towards each other for collision have ultrarelativistic speed and so they are Lorentz contracted. This implies that initially there is a spatial anisotropy in the system. As the system needs to thermalize and isotropize, it generates an anisotropic pressure gradient in all directions. In this way many processes are involved including the formation of unstable modes. These interactions map the spatial anisotropy to momentum anisotropy which remains throughout the evolution of such a system. So, it is important to include momentum anisotropy in the analysis [25]. The effect due to a local anisotropy of the momentum space on the heavy-quark potential is studied in a previous work [40]. The anisotropy is caused by external fields in studying the properties of quarkonium states [41–46]. Besides, the anisotropic quark-gluon plasma concepts are reviewed in [47]. The phenomenological studies that can reproduce the experimentally measured $R_{AA}$ of bottomonia with and without recombination are given in [48, 49], respectively.

So far, most of the previous calculations concentrate on the study of the heavy bound-state properties at vanishing baryonic chemical potential of the thermal medium as in [25, 26, 46]. There are little works which deal with a nonvanishing chemical potential in both perturbative and nonperturbative approaches. The color-screening effect at finite temperature and chemical potential was studied in a thermofield dynamics approach [50, 51] in which the phenomenological potential model [52] and an error-function-type confining force with a color-screened Coulomb-type potential were used. Lattice QCD has also been used to study the color screening in the heavy-quark potential at finite density with Wilson fermions [53]. Kakade and Patra [54] have investigated the quarkonium dissociation at lower temperatures and higher baryonic chemical potential by correcting both the perturbative and nonperturbative approaches of the Cornell potential through the dielectric permittivity in an isotropic medium. The equation of state has successfully been modified to finite baryonic chemical potential using Taylor expansions [55, 56] around a vanishing chemical potential as well as reweighting techniques [57, 58] and using imaginary chemical potentials [59]. The holographic description of the thermal behavior of $b\overline{b}$ heavy vector mesons inside a plasma at finite temperature and density is studied in [60].

The exploration of high baryon densities and the moderate-temperature region of the QCD phase diagram is possible with the upcoming compressed baryonic matter (CBM) experiment at the Facility for Antiproton and Ion Research (FAIR) [54]. Moreover, model calculations based on transport and hydrodynamical equations show that the highest net baryon densities ($\rho_B$) carried out in the center of collision are ~6 to 12 times the density of normal nuclear matter for the most central collision ($b = 0$) [61].

The aim of this work is to study quarkonium dissociation of the quark matter at finite temperature and baryonic chemical potential in an anisotropic medium in comparison with an isotropic medium, in which the N-dimensional Schrödinger equation is analytically solved using the Nikiforov-Uvarov (NU) method. In Section 2, the multidimensional Schrödinger equation is introduced with heavy-quark potential in an anisotropic medium. Moreover, the solution of the N-dimensional Schrödinger equation is given by using the NU method. In Section 3, the binding energy and dissociation temperature are studied with different parameters in the three-dimensional space. In Section 4, the summary and conclusion are presented.

## 2. The N-Dimensional Schrödinger Equation at Finite Temperature in an Anisotropic Medium and Baryonic Chemical Potential

The Schrödinger equation for two particles interacting via an effective potential in the $N$-dimensional space is given [14, 62–64]:

$$\left[ \frac{d^2}{dr^2} + \frac{(N-1)}{r} \frac{d}{dr} - \frac{l(l+N-2)}{r^2} + 2\mu (E - V(r)) \right] \tag{1}$$
$$\cdot \Psi (r) = 0,$$

where $\mu, l$, and $N$ are the reduced mass for the quarkonium particle, the angular momentum quantum number, and the dimensionality number, respectively. Setting the wave function $\Psi(r) = r^{(1-N)/2} R(r)$, the following radial Schrödinger equation is obtained

$$\left[ \frac{d^2}{dr^2} + 2\mu \left( E - V(r) - \frac{(l+(N-2)/2)^2 - 1/4}{2\mu r^2} \right) \right] \tag{2}$$
$$\cdot R(r) = 0.$$

*2.1. Real Part of the Potential in a Anisotropic Medium.* Here, we aim to find the potential due to the presence of a dissipative anisotropic hot QCD medium. The in-medium modification can be obtained in the Fourier space by dividing the heavy-quark potential by the medium dielectric permittivity, $\epsilon(K)$, as follows:

$$\widetilde{V}(k) = \frac{V(k)}{\epsilon(K)}, \tag{3}$$

and by taking the inverse Fourier transform, the modified potential is obtained as follows:

$$V(r) = \int \frac{d^3k}{(2\pi)^{3/2}} \left( e^{ik \cdot r} - 1 \right) \widetilde{V}(k), \tag{4}$$

where $V(k)$ is the Fourier transform of Cornell potential $V(r) = -\alpha/r + \sigma r$ that gives

$$V(k) = -\sqrt{\frac{2}{\pi}} \left( \frac{\alpha}{k^2} + \frac{2\sigma}{k^4} \right), \tag{5}$$

$\epsilon(K)$ may be calculated found from the self-energy using finite temperature QCD. By applying hard thermal loop resummation technique as in [25, 46], the static gluon propagator which represents the inelastic scattering of an off-shell gluon to a thermal gluon is defined as follows:

$$\Delta^{\mu\nu}(w, k) = k^2 g^{\mu\nu} - k^\mu k^\nu + \Pi^{\mu\nu}(w, k). \tag{6}$$

The dielectric tensor can be obtained in the static limit in Fourier space, from the temporal component of the propagator as

$$\epsilon^{-1}(K) = -\lim_{w \to 0} k^2 \Delta^{00}(w, k). \tag{7}$$

To calculate the real part of the interquark potential in the static limit, one can obtain first the temporal component of real part of the retarded propagator in Fourier space at finite temperature and chemical potential as given in [54] as follows:

$$
\begin{aligned}
\mathrm{Re}\left[\Delta_R^{00}\right](w = 0, k) = &-\frac{1}{k^2 + m_D^2(T,\mu)} \\
&- \xi\left(\frac{1}{3\left(k^2 + m_D^2(T,\mu)\right)}\right) \\
&- \frac{m_D^2(T,\mu)(3\cos 2\theta - 1)}{6\left(k^2 + m_D^2(T,\mu)\right)^2},
\end{aligned} \tag{8}
$$

The medium dielectric permittivity $\epsilon(K)$ is then given

$$
\begin{aligned}
\epsilon^{-1}(K) = &\frac{k^2}{k^2 + m_D^2} + k^2\xi\left(\frac{1}{3\left(k^2 + m_D^2\right)}\right) \\
&- \frac{m_D^2(3\cos 2\theta - 1)}{6\left(k^2 + m_D^2\right)^2}
\end{aligned} \tag{9}
$$

Substituting (5) and (9) into (3) and then taking its inverse Fourier transform, we can write the real part of the potential for $rm_D \ll 1$ as follows:

$$
\begin{aligned}
V(r,\xi,T,\mu_b) = &\sigma r\left(1 + \frac{\xi}{3}\right) - \frac{\alpha}{r}\left(1 + \frac{(rm_D)^2}{2}\right) \\
&+ \xi\left(\frac{1}{3}\right. \\
&+ \left.\frac{(rm_D)^2}{16}\left(\frac{1}{3} + \frac{(rm_D)^2}{16}\left(\frac{1}{3} + \cos(2\theta)\right)\right)\right),
\end{aligned} \tag{10}
$$

where $\xi$ is the anisotropic parameter. $T$ and $\mu_b$ are the temperature and the baryonic chemical potential, respectively. In (10), the potential depends on $\theta$ which is the angle between the particle momentum and the direction of anisotropy. We note that the potential in (10) reduces to the Cornell potential for $\xi = 0$ and $m_D = 0$ {For details, see [46]}. In the present work, the Debye mass $D(T,\mu_b)$ is given as in [65, 66] by

$$
D(T,\mu_b) = gT\sqrt{\frac{N_c}{3} + \frac{N_f}{6} + \frac{N_f}{2\pi^2}\left(\frac{\mu_q}{T}\right)^2}, \tag{11}
$$

where $g$ is the coupling constant as defined in [51], $\mu_q$ is the quark chemical potential ($\mu_q = \mu_b/3$), $N_f$ is number of flavours, and $N_c$ is number of colors. The NU method [67] is briefly given here to solve the form of the following:

$$
\Psi''(s) + \frac{\overline{\tau}(s)}{\sigma(s)}\Psi'(s) + \frac{\widetilde{\sigma}(s)}{\sigma^2(s)}\Psi(s) = 0, \tag{12}
$$

where $\sigma(s)$ and $\widetilde{\sigma}(s)$ are polynomials of maximum second degree and $\overline{\tau}(s)$ is a polynomial of maximum first degree with an appropriate $s = s(r)$ coordinate transformation. We try to find a particular solution by separation of variables, if one deals with the transformation

$$
\Psi(s) = \Phi(s)\chi(s). \tag{13}
$$

Equation (13) is written as

$$
\sigma(s)\chi''(s) + \tau(s)\chi(s) + \lambda\chi(s) = 0, \tag{14}
$$

where

$$
\sigma(s) = \pi(s)\frac{\Phi(s)}{\Phi'(s)}, \tag{15}
$$

$$
\tau(s) = \overline{\tau}(s) + 2\pi(s); \quad \tau'(s) < 0, \tag{16}
$$

$$
\lambda = \lambda_n = -n\tau'(s) - \frac{n(n-1)}{2}\sigma''(s), \tag{17}
$$
$$
n = 0, 1, 2, \ldots,
$$

$\chi(s) = \chi_n(s)$ is a polynomial of degree $n$ which satisfies the hypergeometric equation, taking the form

$$
\chi_n(s) = \frac{B_n}{\rho_n}\frac{d^n}{ds^n}\left(\sigma''(s)\rho(s)\right), \tag{18}
$$

where $B_n$ is a normalization constant and $\rho(s)$ is a weight function which satisfies the following:

$$
\frac{d}{ds}\omega(s) = \frac{\tau(s)}{\sigma(s)}\omega(s); \quad \omega(s) = \sigma(s)\rho(s), \tag{19}
$$

$$
\begin{aligned}
\pi(s) = &\frac{\sigma'(s) - \overline{\tau}(s)}{2} \\
&\pm \sqrt{\left(\frac{\sigma'(s) - \overline{\tau}(s)}{2}\right)^2 - \widetilde{\sigma}(s) + K\sigma(s)},
\end{aligned} \tag{20}
$$

$$
\lambda = K + \pi'(s), \tag{21}
$$

$\pi(s)$ is a polynomial of the first degree. The values of $K$ in the square root of (21) are possible to calculate if the function under the square is a square of a function. This is possible if its discriminant is zero. For $r$ parallel to the direction of $n$ of anisotropy at $\theta = 0$, the potential is given by

$$
V(r) = a_1 r - \frac{b_1}{r}, \tag{22}
$$

where

$$
a_1 = \sigma + \frac{1}{3}\sigma\xi - \frac{1}{2}\alpha m_D^2 - \frac{1}{2}\alpha\xi m_D^2, \tag{23}
$$

$$
b_1 = \alpha + \frac{\alpha\xi}{3}. \tag{24}
$$

By applying the above method to the potential given in (22), we obtain the energy eigenvalues as follows:

$$
\begin{aligned}
E_{nl}^\parallel = &\frac{3a_1}{\delta} \\
&- \frac{2\mu\left(3a_1/\delta^2 + b_1\right)^2}{\left[(2n+1) + 8\mu a_1/\delta^3 + 4\left((l + (N-2)/2)^2 - 1/4\right)\right]^2},
\end{aligned} \tag{25}
$$

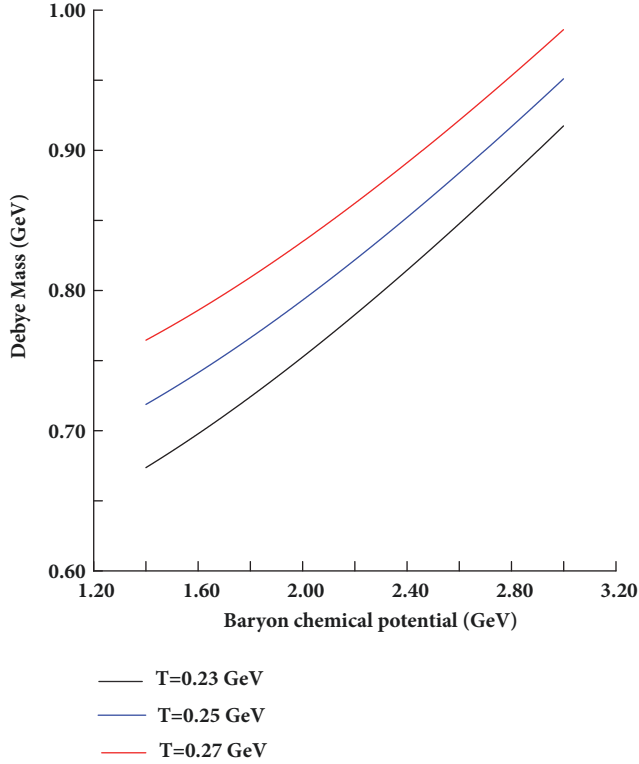

FIGURE 1: Debye mass is plotted as a function of baryonic chemical potential at different values of temperature.

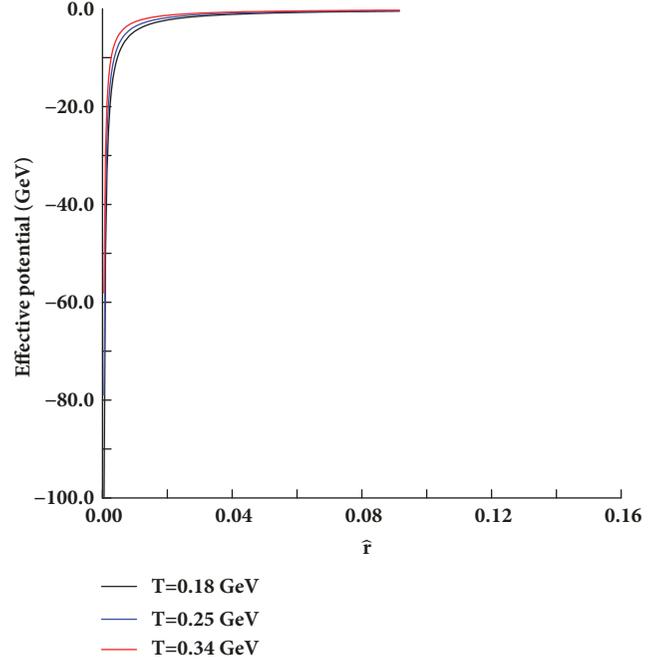

FIGURE 2: The potential divided by $(g^2 C_f r m_d)$ is plotted as a function of $\hat{r}$ at zero chemical potential and $N = 3$ for different values of temperatures.

Similarly, for $r$ perpendicular to the direction of $n$ of anisotropy at $\theta = \pi/2$, the potential is given by

$$V(r) = a_2 r - \frac{b_2}{r}, \tag{26}$$

where

$$a_2 = \sigma + \frac{1}{3}\sigma\xi - \frac{1}{2}\alpha m_D^2 + \frac{1}{24}\alpha\xi m_D^2, \tag{27}$$

$$b_2 = \alpha + \frac{\alpha\xi}{3}. \tag{28}$$

and the energy eigenvalues are given as follows:

$$E_{nl}^{\perp} = \frac{3a_2}{\delta}$$
$$- \frac{2\mu\left(3a_2/\delta^2 + b_2\right)^2}{\left[(2n+1) + 8\mu a_2/\delta^3 + 4\left((l + (N-2)/2)^2 - 1/4\right)\right]^2}, \tag{29}$$

where $\delta$ is a parameter determined as in [68].

## 3. Discussion of the Results

In the present analysis, the various quantities are computed, using the weakly anisotropy parameter ($\xi \preceq 0$) at finite temperature and finite baryonic chemical potential in a QCD plasma. First, we discuss the Debye mass which plays an important role in the present study, since the Debye mass is inserted through the potential and also in the binding energy. We assume the Debye mass intact from the effects of anisotropy present in the media so that it remains the same in both media (an isotropic and an anisotropic) as done in [43, 46]. In Figure 1, the Debye mass is plotted as a function of the baryonic chemical potential, for three values of temperatures (T = 0.23 GeV), (T = 0.25 GeV), and (T = 0.27 GeV). The Debye mass is an increasing function of the chemical potential. By increasing the temperature, we note that the Debye mass shifts to higher values. Hence, the Debye mass is affected in an isotropic medium when the finite baryonic chemical potential is included. In [54], the authors studied the quarkonium mass at the lower temperatures and higher chemical potentials. They found that the Debye mass increases with increasing the chemical potential which is compatible with Figure 1. In addition, the Debye mass increases with increasing temperatures, where they studied the Debye mass below the range of temperatures 0.02 to and up 0.12 GeV. This conclusion is observed in [50], in which the color-screening effects in the QGP are noticeable at higher temperatures and lower baryon density in the anisotropic medium.

In Figure 2, the effective potential given by (20) is plotted as a function of $\hat{r}$ ($\hat{r} = rm_D$) at different values of temperatures. We note that the potential has negative values for all values of $\hat{r}$. Thus, the effective potential has an attractive character. The potential is a little sensitive to changing the in-medium modification of the anisotropic parameter. Therefore, the potential is attractive for isotropic and anisotropic media. In [25], in the isotropic medium, medium modification to the linear term remains positive up to 2-3 $T_c$, making the potential less attractive. In contrast, in the anisotropic case medium modification to the linear term becomes negative and the overall full potential becomes more attractive. In [40], the potential is attractive for all values of $\hat{r}$ and the potential is sensitive to $\xi$ when $\xi$ is greater than 1. In the present study, we restricted the present calculations to weakly anisotropic parameter $\xi \leq 0$ as in [25, 46]. We found that the potential is a little sensitive to the baryonic chemical potential that is in agreement with [62]. For the real potential at $\theta = \pi/2$, we found that the potential is marginally affected in comparison with the potential at $\theta = 0$. This conclusion is in agreement with [46].

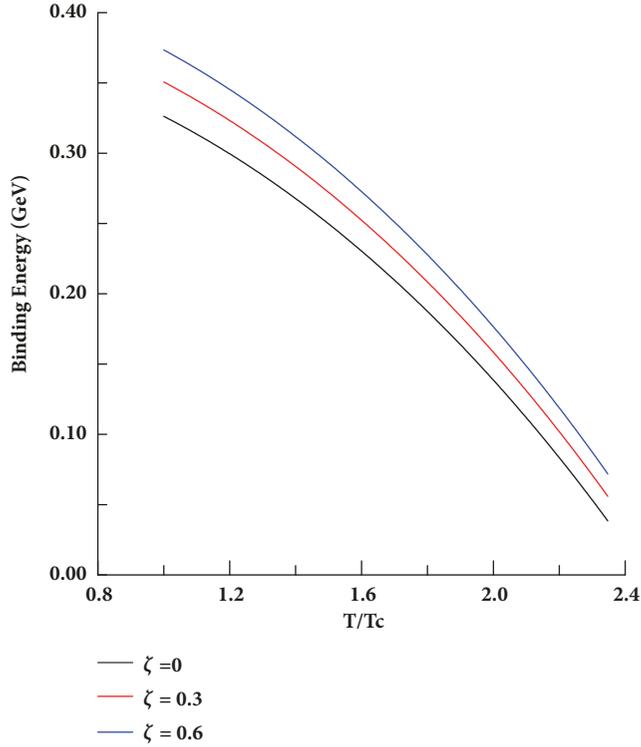

FIGURE 3: The binding energy of 1S charmonium is plotted as a function of temperature at zero chemical potential and N = 3 for different values of $\xi$.

### 3.1. Binding Energy of the Quarkonium.

Here, heavy-quark potential is obtained as a function of temperature and baryonic chemical potential and we solve the N-Schrödinger equation by using NU method in the two cases: when particle momentum is along the direction of anisotropy and when it is perpendicular to the direction of anisotropy. The analytic solution of N-Schrödinger equation gives a good accuracy for the binding energy when $\hat{r} \leq 1$. In the following, we examine the binding energy of quarkonium with variations of the temperature ($T$) and the baryonic chemical potential ($\mu_b$).

In Figure 3, the binding energy of 1S state of charmonium is plotted as a function of temperature at zero baryonic chemical potential. There are basically two features: first, when the anisotropy increases, the binding energy becomes stronger in comparison with their isotropic counterpart since the potential becomes deeper with the increase of anisotropy due to the weaker screening. The screening of the Coulomb and string contribution are less accentuated and hence the quarkonium binding energy becomes larger than in the case of the isotropic medium. Also, there is a strong decreasing trend with the temperature. This is due to the fact that the screening becomes stronger with the increase of the temperature. Our results on the temperature dependence of the binding energies are in agreement with similar works in previous works [25]. In Figure 4, the binding energy is plotted as a function of temperature for different values of the baryonic chemical potential in the isotropic medium ($\xi = 0$). Figure 4 shows that the binding energy decreases with increasing the temperature. The binding energy shows a weak dependence on the baryonic chemical potential up to $\mu_b = 0.6$ GeV for all values of temperatures above $T_c$. This is due to the fact that the screening becomes stronger with increasing baryon chemical potential and therefore the effective potential is weakened in comparison with $\mu_b = 0$ case. Thus, the study of the binding energies at finite temperature and finite baryonic chemical potential gives us indication for the dissociation temperature as we will see in next subsection.

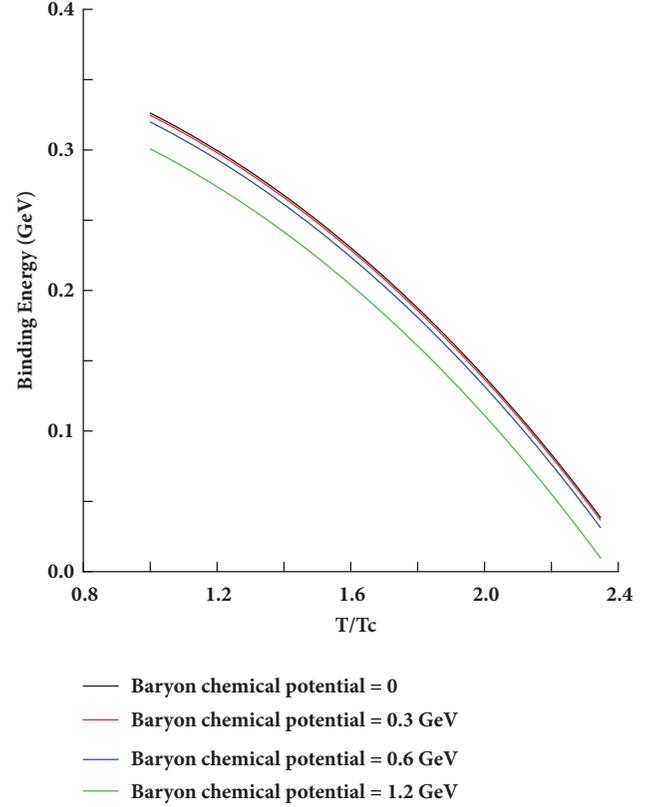

FIGURE 4: The binding energy of 1S charmonium is plotted as a function of temperature at $\xi = 0$ and parameter and $N = 3$ for different values of baryonic chemical potential.

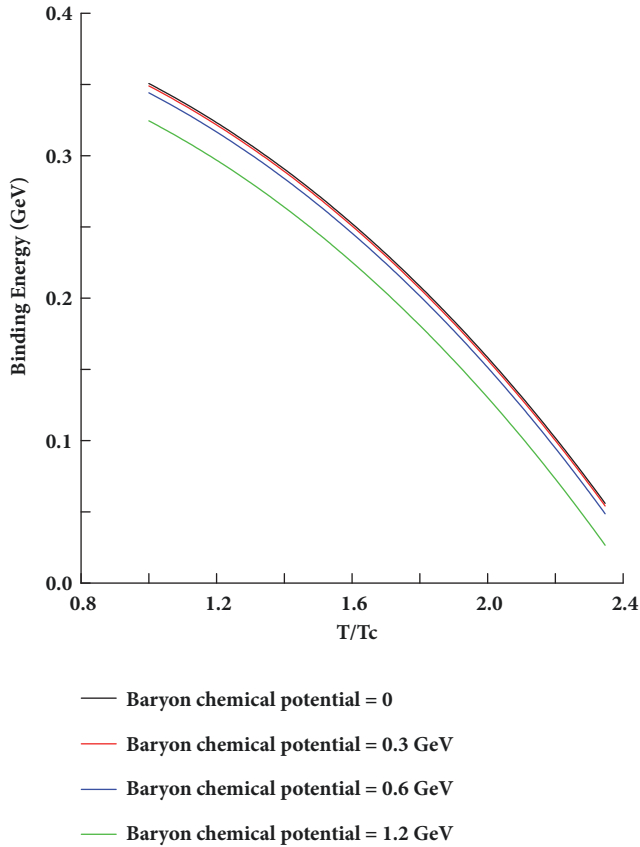

FIGURE 5: The binding energy of 1S charmonium is plotted in temperature at $\xi = 0.3$ and $N = 3$ for different values of baryonic chemical potential.

TABLE 1: Dissociation temperature $(T_D)$ at $\mu_b = 0$ and $N = 3$.

| $T_D$ is in units of $T_c$ | | |
| --- | --- | --- |
| States | $\xi = 0$ | $\xi = 0.3$ |
| $J/\Psi$ | 1.429 | 1.494 |
| Ref. [24] | 1.3 | - |
| Ref. [25] | 1.38 | 1.41 |
| Ref. [26] | 1.40 | 1.46 |
| Ref. [47] | 1.520 | 1.55 |
| $T_D$ is in units of $T_c$ | | |
| States | $\xi = 0$ | $\xi = 0.3$ |
| $\Upsilon$ | 2.201 | 2.276 |
| Ref. [24] | 1.6 | - |
| Ref. [25] | 1.70 | 1.71 |
| Ref. [26] | 3.10 | 3.17 |
| Ref. [47] | 2.964 | 3.062 |

In Figure 5, the binding energy decreases with increasing the temperature for different values of the baryonic potential in the anisotropic medium ($\xi = 0.3$). In addition, the binding energy shows weak dependence on the baryonic chemical potential up to 0.6 GeV. We note that similar behavior is observed for isotropic and anisotropic media. However, the binding energy is deeper in comparison with the isotropic medium in Figure 4. Thus, the bound state is large in the anisotropic medium.

In Figures 6 and 7, the binding energy is plotted in 3D for the isotropic medium ($\xi = 0$) and anisotropic medium ($\xi = 0.3$), respectively. We note that the binding energy decreases with increasing the temperature for any value of the baryonic chemical potential and it has a little change with increasing the baryonic chemical potential. In addition, we note that the binding energy is higher in the anisotropic medium. Therefore, the bound state is enhanced in an anisotropic medium.

### 3.2. Dissociation of Heavy Quarkonia at Finite Temperature and Chemical Potential.

The dissociation of a two-body bound state in a thermal medium can be given qualitatively: when the binding energy of a resonance state drops below the mean-thermal energy of quarkonium, i.e,. the state becomes weakly bound. In [69], the authors concluded that one need not to reach the binding energy $(E_b)$ to be zero for the dissociation but a weaker condition is assumed $E_b \leq T$ which causes a state to be weakly bound. In fact, when $E_b \simeq T$, the resonances have been broadened due to direct thermal activation. So the dissociation of the bound states may be expected to occur roughly around $E_b \simeq T$. Thus, we say that there are two criteria for the dissociation of quarkonia bound state in the QGP medium. The first one is the dissociation of a given quarkonia bound state by the thermal effects alone. On the other hand, the second criterion is based on the dissolution of a given quarkonia state while its thermal width is overcome by twice the value of the real part of the binding energy [26, 46, 70]. In the present analysis, we study the effect of isotropic and anisotropic media on the dissociation temperatures in the presence of hot or dense media for 1S state of charmonium and 1S state of bottomonium. In addition, we compare the present results with recent published works.

In Figure 8, the binding energy is plotted as a function of baryonic chemical potential in the isotropic medium ($\xi = 0$) for 1S state of the charmonium. We note that the binding energy decreases with increasing the baryonic potential for all values of temperatures above the critical temperature ($T_c = 0.17$ GeV). Hence, the dissociation temperature is sensitive to the change of the baryonic chemical potential in the range $\mu_b = 1.2$ GeV to 3 GeV. A similar situation occurs in an anisotropic medium in Figure 9, in which the binding energy is sensitive for changing the baryonic chemical potential in the same range of the baryon chemical potential. In [55], the author calculated the binding energy as a decreasing function of the baryonic chemical potential and the binding energy drops to lower values by increasing the temperature in the isotropic medium. In our present results in Figure 9, we notice an agreement with [55].

In Table 1, the dissociation temperatures $(T_D)$ at $\xi = 0$ are less than in the anisotropic case at $\xi = 0.3$ for 1S of the charmonium and 1S of the bottomonium. This indicates that the bound state is more bound in the anisotropic medium. This conclusion is also noted in [25, 26, 46]. We note that 1S state of $J/\Psi$ is in good agreement with [25, 26, 46]. The dissociation temperature for 1S state of the charmonium is less than the 1s state of the bottomonium in the present work. This conclusion is also noted in [24–26, 46].

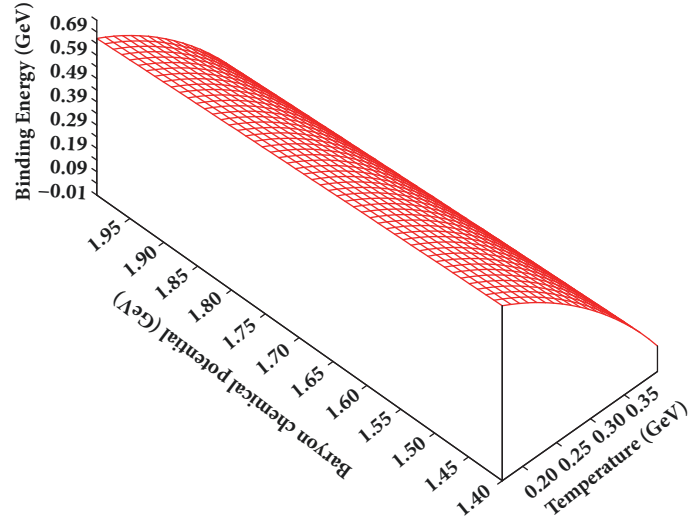

FIGURE 6: The binding energy of 1S charmonium is plotted as a function of temperature and baryonic chemical potential in 3D space in the isotropic medium ($\xi = 0$).

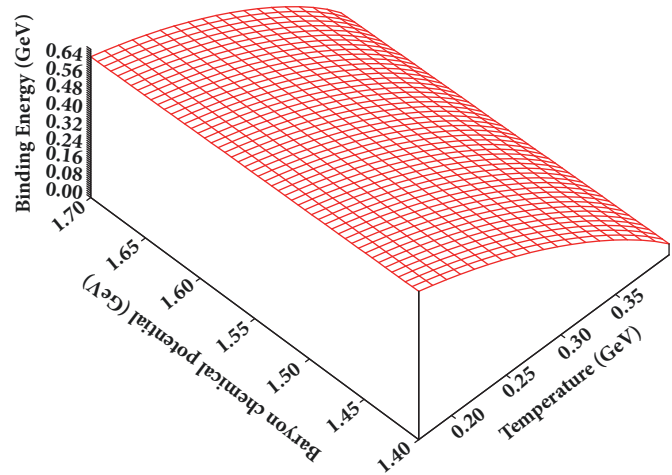

FIGURE 7: The binding energy of 1S charmonium is plotted as a function of temperature and baryonic chemical potential in 3D space in the anisotropic medium ($\xi = 0.3$).

## 4. Summary and Conclusion

In this work, the dissociation temperature for quarkonia such as the 1S charmonium and 1S bottomonium states have been calculated using the medium modified interquark potential in the anisotropic medium when baryonic chemical potential is included. The real part of the heavy-quark potential is discussed in hot and dense QCD media. The binding energy is obtained in an anisotropic medium by solving the N-Schrödinger equation using NU method in the two cases: Debye mass for quark pairs which are aligned along the direction of anisotropy and the transverse alignment. We give the binding energy in the N-dimensional space. Therefore, we can obtain the binding energy in the lower dimensions as used here at $N = 3$. The quarkonium states are studied in the N-dimensional space as in [71].

We observed that the quarkonia states are more bound and the results survive the higher temperature in comparison with the isotropic medium when the finite temperature and zero baryonic chemical potential are included. The current study shows that the baryonic chemical potential has a small effect on the binding energy of quarkonia at higher values of temperatures up to $\mu_b = 0.6$ GeV. The present results show that the dissociation temperature increases with increasing the anisotropic parameter for 1S state of charmonium and bottomonium. The present results are in agreement with previous studies as in [25, 26, 46]. In these studies, the authors calculated the binding energy and dissociation temperatures in the anisotropic medium by using three-dimensional Schrödinger equation, in which the baryonic chemical potential is not considered. Thus, the present study shows the role of the baryonic chemical potential in an anisotropic hot QCD medium. We hope to investigate in a future work the dissociation temperature using the decay width of quarkonium bound states from the imaginary part of the potential in the dense medium with an external magnetic field.

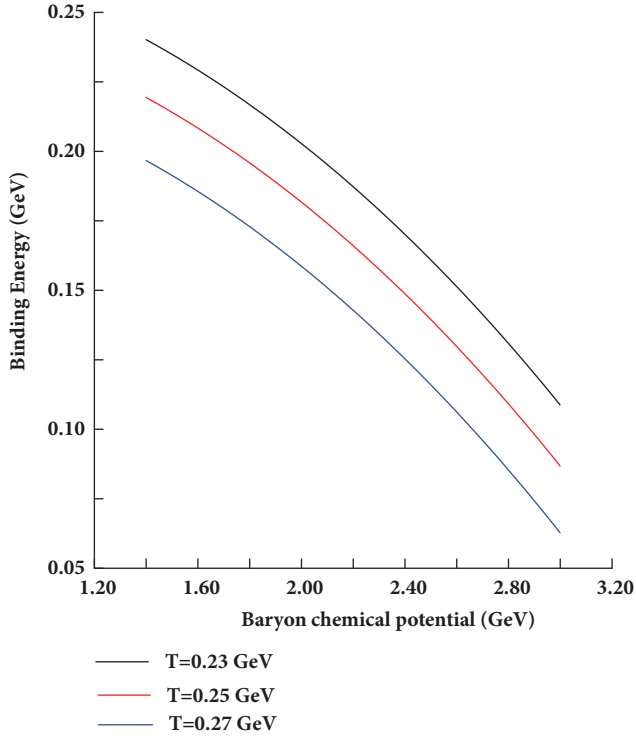

Figure 8: The binding energy 1S charmonium is plotted as a function of baryonic chemical potential for different values of temperature at $\xi = 0$.

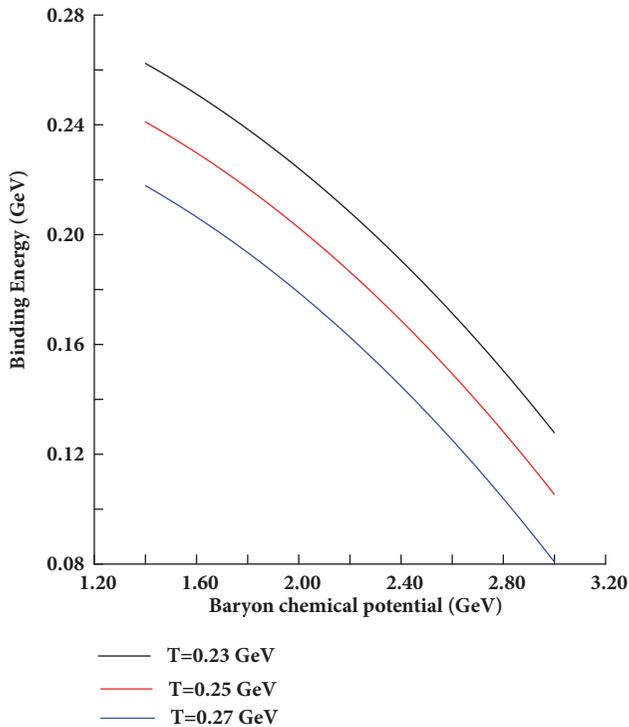

Figure 9: The binding energy 1S charmonium is plotted as a function of baryonic chemical potential for different values of temperature at $\xi = 0.3$.

## Data Availability

The information given in our tables is available for readers in the original references listed in our work.

## Disclosure

H. M. Mansour is Fellow of the Institute of Physics.

## Conflicts of Interest

The authors declare that they have no conflicts of interest.

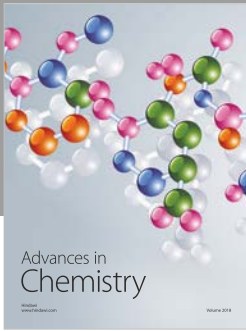

Advances in
**Chemistry**

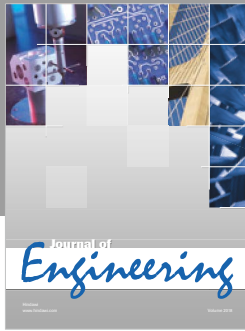

*Journal of*
**Engineering**

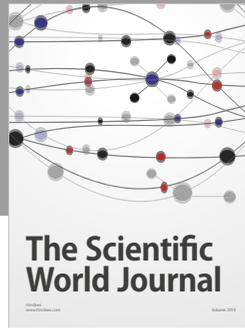

**The Scientific
World Journal**

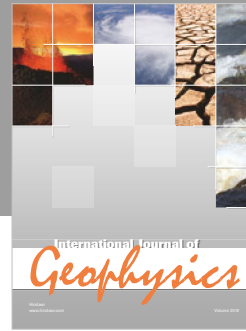

*International Journal of*
**Geophysics**

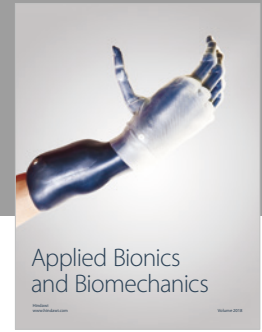

Applied Bionics
and Biomechanics

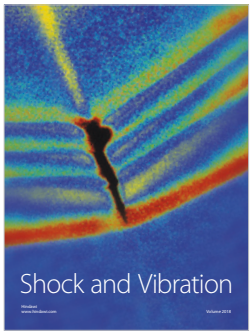

Shock and Vibration

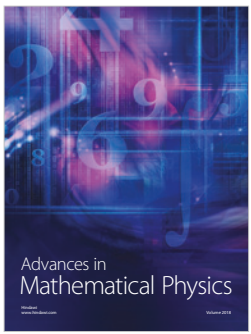

Advances in
Mathematical Physics

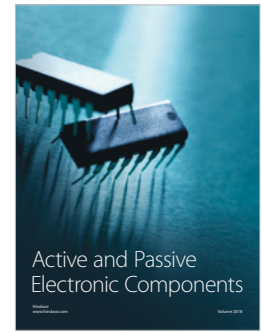

Active and Passive
Electronic Components

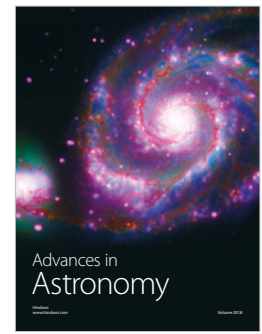

Advances in
Astronomy

**Hindawi**

Submit your manuscripts at
www.hindawi.com

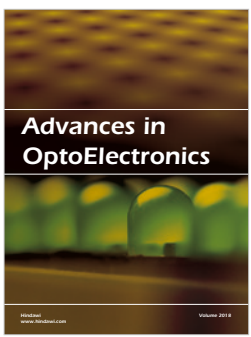

Advances in
**OptoElectronics**

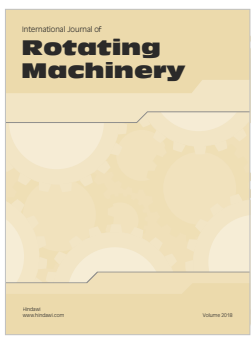

*International Journal of*
**Rotating
Machinery**

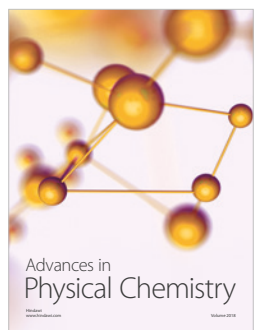

Advances in
**Physical Chemistry**

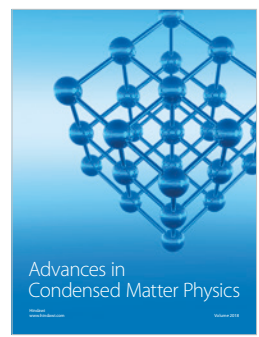

Advances in
Condensed Matter Physics

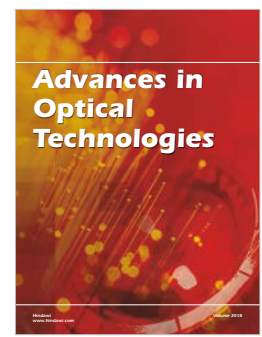

**Advances in
Optical
Technologies**

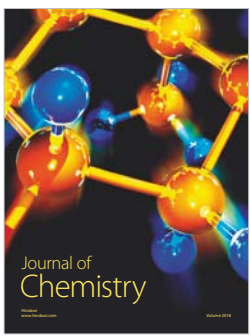

Journal of
**Chemistry**

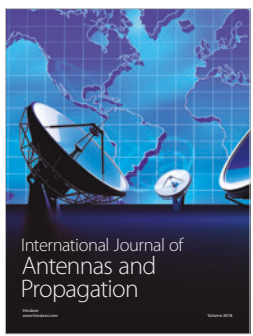

International Journal of
**Antennas and
Propagation**

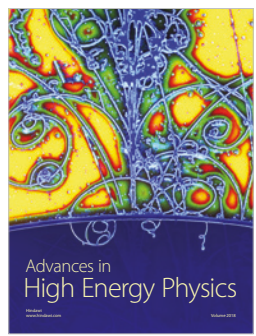

Advances in
**High Energy Physics**

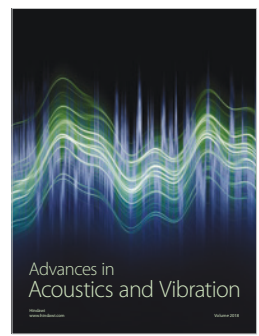

Advances in
**Acoustics and Vibration**

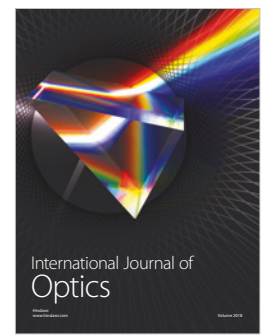

International Journal of
**Optics**